\documentclass[fleqn,twoside]{article}
\usepackage{espcrc2}

\usepackage{graphicx}

\newcommand{\AmS}{{\protect\the\textfont2
  A\kern-.1667em\lower.5ex\hbox{M}\kern-.125emS}}

\title{Charged Lepton Flavour and CP Violations: Theoretical Impact of Present and 
Future Experiments}

\author{I. Masina\address{Centro Studi e Ricerche  ``E. Fermi'',  
               Rome, Italy and INFN, Sezione di Roma, Rome, Italy}
        and 
        C. A. Savoy\address{Service de Physique Th\'eorique, CEA-Saclay,
        Gif-sur-Yvette, France}%
       \thanks{Work supported in part by the RTN European Program 
        HPRN-CT-2000-00148}}
\begin{document}

\begin{abstract}
We shortly review and emphasize how  $\ell ^{j}\rightarrow  \ell ^{i}\,
\gamma $experiments and the searches for lepton e.d.m. are  constraining
New Physics model building.  They are pure signals of new phenomena
around the TeV scale since the SM contributions are definitely negligible.
It is quite remarkable that they also give  effective tests of the  LFV\&CPV
in seesaw couplings and in grand-unified theories. In particular, the limits on
$d_{e}$ nicely complement the proton decay bounds in selecting O(10)
models.
\vspace{1pc}
\end{abstract}

\maketitle

\section{Introduction}

The title of this short review could well be ``Lepton Flavour and CP Violations 
without Neutrinos for Neutrino Physicists'' since the aim here is to emphasize the impact 
of the present and near future experiments on charged lepton transitions on ({\it a}) 
the flavour and CP pattern of the new lepton-like states in theories beyond the SM and  
({\it b}) the neutrino mass models. Indeed, FCNC and CPV are generic problems for
theories that postulate new states at the TeV scales which is typically the case in string 
inspired solutions to the hierarchy problem: squark and sleptons in supersymmetric models
and Kaluza-Klein states in models with large compact dimensions. Basically, the problems
arise because particles get mass from different, generically unrelated, 
mechanisms: Higgs couplings for the usual fermions, supersymmetry breaking or 
compactification for the new heavy states. Any misalignment  in flavour space or in 
CP phases between the resulting mass matrices yields FCNC or CPV effective operators 
from radiative corrections involving the heavy states.  The present experiments already put 
either strong  limits on these misalignments or lower bounds above the TeV for the new 
state masses. These constraints open a new framework to investigate both the flavour
problem and the structure of the new theories, in particular, in the lepton sector.

Let us first summarize the  basic facts.

\noindent - Neutrino oscillations require large lepton flavour (as defined by the charged
leptons) mixing in the effective neutrino mass matrix, but this does not necessarily 
imply large mixing in the charged sector.
 
\noindent - Leptogenesis is viable within the seesaw model for neutrino masses if the 
Yukawa couplings of the neutrinos to the Higgs (\textsl{i.e.,} the Dirac masses)
have sizeable CP phases.

\noindent - In the somewhat minimal framework where only the seesaw mechanism
to generate neutrino masses is added to the Standard Model, the LFV\&CPV
effects in the charged lepton physics are extremely tiny because of the analogue of the
GIM mechanism: they depend on factors of $\Delta m_{\nu}^{2}/ M_{W}^{2} < 
10^{-24}$.

\noindent - Therefore,  if observed in the current or planned experiments, LFV in
 charged decays ($\mu \rightarrow e\,\gamma ,\: \tau \rightarrow \mu \,\gamma $) 
 or CPV electric dipole moments ($d_{e}, \, d_{\mu }$) would be signals of 
 New Physics beyond the SM and seesaw models around the TeV region (\textit{e.g.,}
 supersymmetry).

\noindent - Conversely, the present and future bounds on these LFV\&CPV transitions 
constrain New Physics around the TeV region (\textit{e.g.,} supersymmetry) and 
require a LFV\&CPV inhibition mechanism in the corresponding theories. 

\noindent - Some of these experiments provide already relevant constraints on radiative 
corrections from new LFV\&CPV couplings in theories at unattainable scales: GUT's, 
seesaw, flavour models ...

We shall emphasize here the last point to demonstrate the importance of these experiments.
Indeed they provide already restrictions on the masses and couplings of  very heavy states 
in seesaw and/or GUT models from their virtual contributions in radiative corrections
to the effective low energy parameters in any new theory in the TeV region 
(\textit{e.g.,} slepton masses in supersymmetry). Future experiments will make these
constraints even more impressive. 

The phenomenological information gathered from LFV\&CPV in the charged sector is
quite complementary to those provided by measurements of neutrino oscillations or 
from the assumption of leptogenesis for the generation of the baryon asymmetry in 
the universe. This has been discussed in several papers and reviews at a more technical
level \cite{davidson} and, here, we shall rather illustrate this aspect by 
comparing the explicit predictions of simple basic models with the available 
set of experimental data.

The variety of these data are reviewed by M. Aoki in these Proceedings \cite{aoki},
which we refer to for more detailed description of different experiments and prospects 
and for the relevant bibliography. It is enough for our discussion to concentrate on 
the more relevant ones and to just keep the orders of magnitude of the experimental 
bounds, as shown in Table~\ref{table:1}. Notice that electric dipole moments are 
displayed in units of fm rather than cm - this is kind of more natural in comparing 
with the anomalous magnetic moments. 
\begin{table*}[htb]
\caption{Orders of magnitude of present experimental bounds on LFV\&CPV 
and planned improvements}
\label{table:1}
\newcommand{\m}{\hphantom{$-$}}
\newcommand{\cc}[1]{\multicolumn{1}{c}{#1}}
\renewcommand{\tabcolsep}{2pc} 
\renewcommand{\arraystretch}{1.2} 
\begin{tabular}{@{}lllll}
\hline
\textsc{observable} & \cc{\textsc{present limit}} & \cc{\textsc{prospects}} & 
\cc{\textsc{SM prediction}} \\
\hline
CLFV & B.R. & B.R. & B.R.  \\
\hline
$\tau \rightarrow \mu \,\gamma $    & $10^{-6}$ & $10^{-8}$ & $10^{-48}$ \\
$\mu \rightarrow e\,\gamma$  &$10^{-11}$ &$10^{-14}$  & $10^{-48}$  \\
\hline
EDM   &  e.fm &  e.fm  &  e.fm \\
\hline
$d_{\mu }$ & $10^{-5}$ & $10^{-11}$ & $10^{-22}$ &  \\
$ d_{e} $ & $10^{-14}$ & $10^{-16}$ & $10^{-25}$ \\
\hline
\end{tabular}\\[2pt]
For the detailed results and references, see Ref. \cite{aoki}, in these Proceedings.
\end{table*}

\section{Impact on New Physics at O(TeV) Scales}

In order to evaluate to which extent these data constrain theories beyond the
SM (notice the SM predictions in the table) first notice that all these transitions
correspond to the same family of operators as the anomalous magnetic moment.
Therefore they can all be grouped in the expression:
\begin{equation}
\mu _{ij} +id_{ij} = \frac{\Gamma ^{NP}_{ij}}{M^{2}_{NP}}\, 
\frac{e\,m_{\ell}}{4\pi ^{2}} \: \ell ^{i}_{L}\sigma _{\mu \nu } 
\ell ^{j}_{R} F^{\mu \nu }
\end{equation}
where $M^{2}_{NP}$ and $\Gamma ^{NP}_{ij}$ are a typical heavy mass 
in the radiative loops and  the resulting effective coupling. The lepton mass 
factor, $m_{\ell}$ has been inserted because these couplings are associated
to a $\Delta I = 1/2$ helicity flip  and so require a Higgs v.e.v. and coupling.
The  $4\pi ^{2}$ in the denominator roughly accounts for one-loop factors -
barring more exotic origins for LFV\&CPV. The e.m. transitions gathered in this
expression are: $(g-2)_{\ell ^{i}}$ for $i=j$, the LFV decays $\ell ^{j}
\rightarrow  \ell ^{i}\,\gamma $ for $j>i$, and the CPV e.d.m., 
$d_{\ell ^{i}}$, corresponds to $\mathrm{Im} \Gamma ^{NP}_{ii}$. Let us now 
express $M_{NP}$ in TeV and apply the present bounds in Table~\ref{table:1}.
This is displayed in Table~\ref{table:2} where the limits on $M_{NP}$
are shown in the second column for the present data and the expected improvement 
in these bounds are shown in the third column. Also shown are the naive 
scaling factors between the same e.m. transition for different leptons as given
by a simple choice of the $m_{\ell}$.

\begin{table*}[htb]
\caption{Limits on New Physics contributions on charged lepton LFV\&CPV: 
present limits on the effective scale, expected improvement factor and relative naive
relations}
\label{table:2}
\newcommand{\m}{\hphantom{$-$}}
\newcommand{\cc}[1]{\multicolumn{1}{c}{#1}}
\renewcommand{\tabcolsep}{2pc} 
\renewcommand{\arraystretch}{1.2} 
\begin{tabular}{@{}lllll}
\hline
\textsc{experiment} & $M^{2}_{NP}$ ($\mathrm{TeV} ^{2}$) 
& {\textsc{prospects}} & 
\cc{\textsc{naive scaling}} \\
\hline
$(g-2)_{e}$ & $>\Gamma ^{NP}_{ee} / 1000$  & {} &  
$\propto m^{2}_{e} $ \\ 
$(g-2)_{\mu}$ & $> \Gamma ^{NP}_{\mu \mu } / 20$  & {} & 
$ \propto m^{2}_{\mu }  $  \\
\hline
$\mu \rightarrow e\,\gamma$  & 
$>\Gamma ^{NP}_{\mu e}\times 20$ & $\times 30$ & 
$\propto m_{\mu} $ \\
$\tau \rightarrow \mu \,\gamma $  & 
$>\Gamma ^{NP}_{\tau \mu } / 40$ & $\times 10$ & 
$\propto 0.2m_{\tau }$ \\
\hline
$ d_{e} $ & $>\mathrm{Im} \Gamma ^{NP}_{e e}\times 70$ & 
$\times 100 $ & $\propto m_{e}$ \\
$d_{\mu }$ & $ >\mathrm{Im} \Gamma ^{NP}_{\mu \mu}\times 10^{-5}$ 
& $\times 10^{6}$ & $\propto m_{\mu}$   \\
\hline
\end{tabular}\\[2pt]
\end{table*}
The value of the new mass scale $M_{NP}$ can be as low as O($200$GeV) without
violating the limits on $(g-2)_{\mu}$. On the contrary, bounds on $d_{e}$ already
require a suppression of two orders of magnitude in the CP violating effective coupling 
if $M_{NP}=$ O(TeV). 

It is a well known fact that new physics is expected when experiments will go beyond
the present TeV frontier in order to explain the success of the SM as a very predictive
effective theory. Even on general grounds, the experimental data on LFV\&CPV
are already providing relevant restrictions on the flavour pattern of this new physics.

\subsection{Constraints on supersymmetric models}

\begin{figure}[htb] 
\centerline{ 
\resizebox*{8cm}{!}{\includegraphics{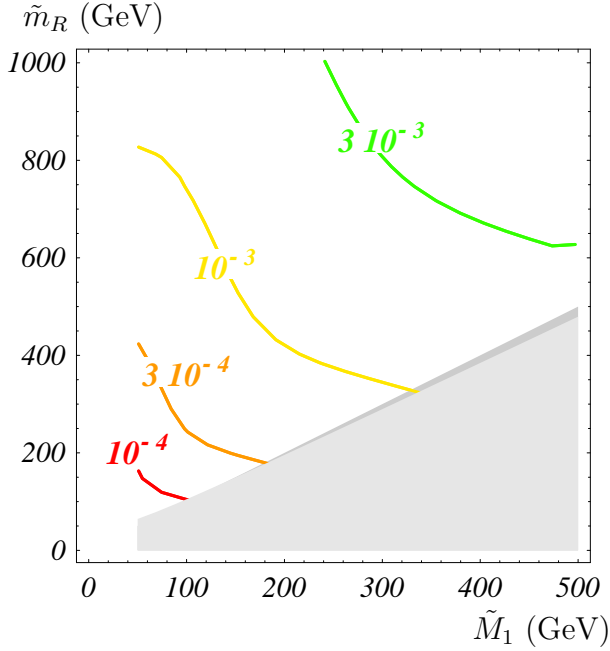}}
\put(-50,-5){\large ${\tilde M_1}$ (GeV)}
\put(-220,227){\large $\tilde m_R$ (GeV)} }
\caption{Upper bound on $|\delta^{LL}_{12}|$ for $\tan\beta 
= 10$ and the present limit BR$(\mu\rightarrow e \gamma) \le 
10^{-11}$.  From \cite{sleptonarium}. }
\label{fig:1} 
\end{figure}

\begin{figure}[htb] 
\centerline{ 
\resizebox*{8.5cm}{!}{\includegraphics{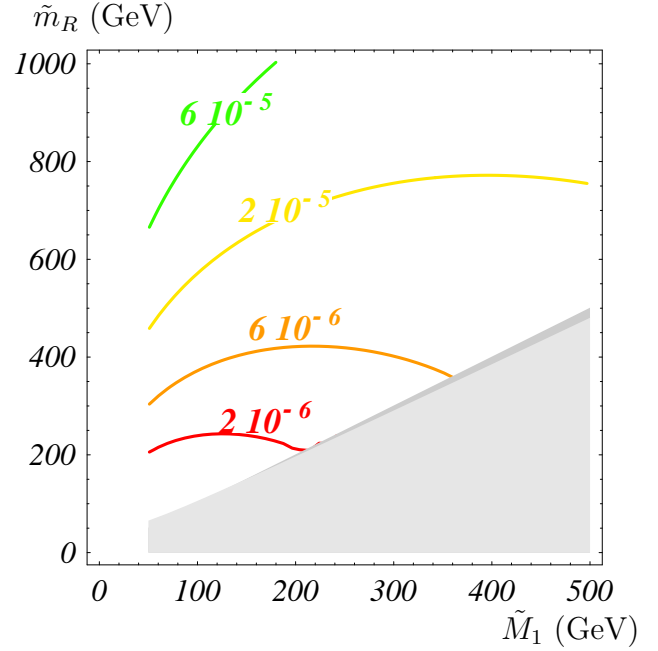}}
\put(-50,3){\large $\tilde M_1$ (GeV)} 
\put(-227,235){\large $\tilde m_R$ (GeV)} \ 
} 
\caption{Upper bound on $|Im(\delta^{LL}_{13} 
\delta^{RR}_{31})|$ for $\tan\beta = 10$ 
and the present limit $ d_e \le 10^{-14}$e fm. 
From \cite{sleptonarium}.} 
\label{fig:2} 
\end{figure}

Supersymmetry is one of the best candidates for the new physics framework and 
it has many sources of LFV\&CPV, in particular, in the slepton mass matrices.
It is worth looking in more detail for the experimental restrictions in the 
case of supersymmetry since we expect the sparticle masses to remain below 
the TeV scale to avoid excessive fine-tuning in the model. Let us consider 
mSUGRA models where, by assumption, the slepton masses have no flavour 
structure and no relevant CP violating phases at the tree-level. We denote 
$\tilde{m}_{L}, \,\tilde{m}_{R}$ the masses of the scalar leptons associated to 
the $(\ell ^{i}_{L}, \,\nu ^{i}_{L})$ doublets and the $\ell ^{i}_{R}$ 
singlet respectively and we skip the so-called $A-$terms for simplicity. 
In order to estimate the allowed deviations in the alignment in flavour space
between $\tilde{m}_{L}, \,\tilde{m}_{R}$ and the charged lepton 
mass matrix, $m_{\ell}$, one usually defines the ratios 
$\delta ^{LL}_{i j}= \tilde{m}^{2}_{L,i j} / \tilde{m}^{2}_{L},\,
\delta ^{RR}_{i j}=\tilde{m}^{2}_{R,i j} /\tilde{m}^{2}_{R},$ with
 $i\neq j,$ and determine the limits on these misalignment ratios as a function of the
 sfermion and slepton masses from the contributions they induce in LFV decays, and
 in CPV e.d.m. if they are complex \cite{gabbiani,sleptonarium}. 
 Two examples of these limits are shown in Figures~\ref{fig:1} and \ref{fig:2} 
 \cite{sleptonarium} in terms of three most relevant parameters: the slepton
 mass $\tilde{m}_{R}$, the gaugino (bino) mass $\tilde{M}_1$ and the Higgs 
 v.e.v. angle, $\tan \beta $.
 
The plots show the allowed flavour misalignment $\delta ^{LL}_{12}$ 
for sparticle masses below $1\,$TeV to be already O($10^{-3}$) -- it will 
become 30 times smaller with the planned experiments -- for the case of 
$\mu \rightarrow e\,\gamma$, actually a stronger limit than the generic
order of magnitude in Table~\ref{table:2} (instead the limits on 
$\delta ^{RR}$ are much more model dependent due to destructive 
interference between different contributions!).  This is the lepton counterpart 
of the supersymmetric flavour problem: the supersymmetric scalar mass generation
must be endowed with such a small misalignment between slepton and lepton 
masses. However, since the orders of magnitude are close to those expected in 
radiative corrections, the alignment must be preserved by quantum contributions. 
This fact will be exploited in the next section.

The limits on the phases in slepton mass matrices obtained from the 
experimental bounds on $d_{e}$ are even stronger as seen in 
Figure~\ref{fig:2} for the product $\mathrm{Im}(\delta ^{LL}_{13} 
\delta ^{RR}_{31}) $ . Indeed, for large $\tan \beta $ their 
imaginary parts are at most of O($10^{-5}$), to go  down  to $10^{-7}$ 
in the next future! Also notice (\textit{cf.} Table~\ref{table:2}) that 
the future measurement of $d_{\mu }$ will provide limits comparable with 
the ones presently extracted from $d_{e}$. It is a remarkable fact that the 
e.d.m. experiments are reaching the accuracy needed to put constraints on 
new physics around the TeV scale at the level of radiative corrections. 
We now turn to discuss how this fact may tell us of properties of theories 
at a scale much above the TeV region, such as the seesaw models and GUT's. 

\section{Constraints on Seesaw and GUT's }

\begin{figure}[htb] 
\centerline{ 
\resizebox*{9cm}{!}{\includegraphics{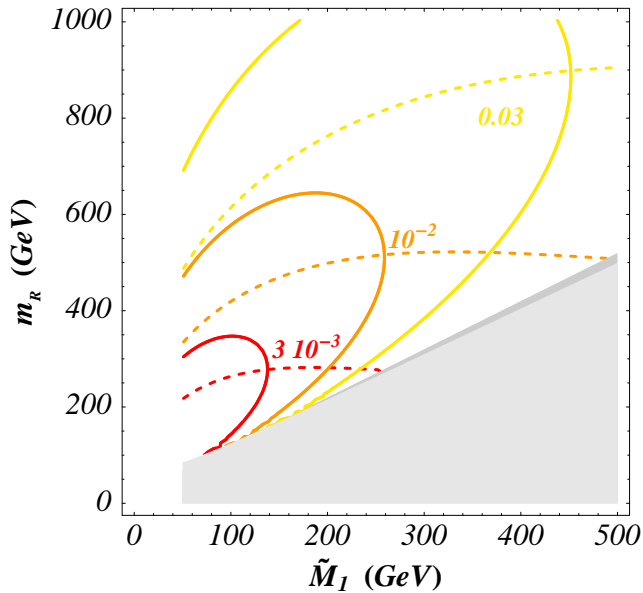}}}
\caption{Upper limit on $|Im(e^{-i \phi_{td}} 
(Y^{\dag }_{\nu }Y_{\nu })_{13})|$ from the present bound on $d_e$, 
for $\tan \beta =10$ and a triplet mass $M_T = 2 \times 10^{16}$ GeV,
$A_0^2=2 m_0^2$ (solid line),  $A_0^2= \tilde M_{1/2}^2$ (dashed line). 
From \cite{isa}.}
\label{fig:3} 
\end{figure}

Very heavy states that couple to sleptons leave their traces in the slepton 
mass matrices through their radiative corrections until they decouple from 
the effective supersymmetric theory. Thus, in the (type I) seesaw model, 
the heavy singlet neutrinos are given very large masses, $M_{R}$, and 
couple to the Higgs and the light leptons and their sleptons through a Yukawa 
matrix $Y_{\nu }$. Correspondingly, they produce a correction to the LFV 
part of $\tilde{m}_{L}$ with a misalignment $\delta ^{LL}_{i j}=
O(1/6) \, (Y^{\dag }_{\nu }\, \ln (M_{P\ell }/M_{R}) \, Y_{\nu })_{ij}$,
where the Planck mass $M_{P\ell }$ is the  cut-off for effective supergravity.
This misalignment is bounded by the LFV decays  \cite{masiborzu}, 
but only the bound on $\delta ^{LL}_{12}$ is really relevant. The phases 
generated only by the seesaw radiative loops could produce lepton e.d.m. at
observable rates only if the  $M_{R}$ eigenvalues are strongly hierarchical 
\cite{ellis,isa}. On the other hand,  in the simplest versions of SU(5) GUT's, 
the colour triplet partners of the Higgs bosons, must get  large masses, 
$M_{T}$, through the doublet-triplet mechanism, while they couple to 
the right-handed leptons through a Yukawa matrix which is approximately $Y_{u}$, 
the usual couplings of the up quarks. Correspondingly, radiative corrections 
are generated to $\delta ^{RR}_{ij}= O(1/6) \, (Y^{\dag }_{u }\, 
\ln (M_{P\ell }/M_{T}) \, Y_{u })_{ij}$. They induce and are constrained 
by  LFV decays as well \cite{barbhall}. What about CPV corrections
to the slepton masses from these heavy neutrinos and colour triplets? It has been 
recently shown \cite{stro,isa} that with SU(5) GUT $\oplus $ seesaw one 
obtains interesting constraints on $Y_{\nu }$ matrix elements from the $d_{e}$ 
experiments, This is shown in  Figure~\ref{fig:3} \cite{isa} for
$(Y^{\dag }_{\nu }Y_{\nu })_{13}$ , where $\phi _{td}$ is the phase in the 
quark CKM matrix. Remember that these limits will be improved by two orders 
of magnitude.
 
Another way to see the relevance of the restrictions from $d_{e}$ on new 
physics is to show that  they are competitive with the well-known constraints 
on the colour triplets masses from proton life-time, $\tau _{p}$ \cite{triplet}. 
Let us take for definiteness $M_{R}\simeq M_{T}\simeq 10^{17}
\mathrm{GeV}$ and  ``typical'' sparticle masses, $\tilde{m}_{R}= 2 \tilde{M}_1=
400\mathrm{GeV}$ and, \textsl{e.g.}, ${\tan \beta = 3}$, with mSUGRA mass
conditions and  let us calculate $d_{e}$ and $\tau _{p}$ in the simplest models
(which are not meant to be more theoretically justified, see \textsl{e.g.} 
\cite{senja}):
(I) SU(5) with seesaw and the simplifying assumption, $Y_{\nu } \approx  Y_{u}$; 
(II) SO(10) with two colour triplets and  $M_{T_{1}} \approx M_{T_{2}}$; 
(III) SO(10) with pseudo-Dirac like colour triplets, $ (r+1)M_{T_{1}} =
(r-1)M_{T_{2}}$ with $r\ll 1 $. The results are summarized in Table~\ref{table:3}.
\begin{table*}[htb]
\caption{\textbf{Electron e.d.m. versus proton lifetime as tests of GUT's.}  
From \cite{triplet}.}
\label{table:3}
\newcommand{\m}{\hphantom{$-$}}
\newcommand{\cc}[1]{\multicolumn{1}{c}{#1}}
\renewcommand{\tabcolsep}{2pc} 
\renewcommand{\arraystretch}{1.2} 
\begin{tabular}{@{}lllll}
\hline
 {} & $ \tau (p\rightarrow K^{+}\nu $)/ ($10^{33}\mathrm{yrs})$ &
  $d_{e}/ (10^{-14}\mathrm{e.fm})$ \\
 \hline
 {\textsc{ experimental bounds}} & $>2$ & $<1$ \\
 
 (I) SU(5) w/ $Y_{\nu }=Y_{u}$ & $O(0.1)$ & $0.2$ \\
  
 (II) SO(10) w/ $M_{T_{1}}=M_{T_{2}}$ & $O(0.1)$ & $2.0$\\
 
 (III) SO(10) w/ $M_{T_{1}}\approx -M_{T_{2}} $ & $0.2\leftrightarrow 7$ &
 $1.8\leftrightarrow 2.0$ \\
 \hline
{}
\end{tabular}\\[2pt]
\end{table*}
While models I and II predict $\tau _{p}$ much below the experimental
value, the pseudo-Dirac trick introduces large negative interference between
the two massive colour triplets increasing $\tau _{p}$ by almost two orders
of magnitude (depending on the various Yukawa phases; note that the quark 
CKM phase in $Y_{u}$ alone is enough unless it is compensated by the 
unknown phases). But the supersymmetric SO(10) models violate the limits 
on $d_{e}$, at least in the simpler versions. While $\tau _{p}$ could be 
reduced by interference, the different contributions add up in the wave function 
renormalisation  that leads to $d_{e}$.

\section{Constraints on Abelian Flavour Models}

In order to explicitly see the impact of the $\mu \rightarrow e\,\gamma$ and
$d_{e}$ experiments on flavour models to explain the fermion masses, including
the neutrinos, let us consider the Froggatt-Nielsen simplest model with an
$U(1)$ flavour group \cite{buch}. It is assumed to be broken by field whose 
v.e.v. defines a small number $\epsilon $ with respect to the cut-off scale of 
the model. Charges are associated to each lepton field as follows: $\ell _{i}$ 
to the doublets $(e,\nu )_{i} $, $e _{i}$ to $e^{c} _{i}$, $n _{i}$ to 
$\nu ^{c} _{i}$, with $i=1,2,3$ as the flavour index, $h_{1,2}$ to the 
two Higgses. In this way one obtains, after the $U(1)$ breaking, effective 
Yukawa couplings endowed with some kind of hierarchy in the eigenvalues 
and the mixings in terms of powers of $\epsilon $ defined by the flavour 
charges. Ordering the heavy right-handed neutrinos by $M_{1}\leq M_{2}
\leq  M_{3}$, the present neutrino oscillation data (mass differences and 
mixing angles for atmospheric and solar neutrinos) basically fix the charges:
$\ell _{3} = \ell _{2} = \ell _{1} -1$, together with $\epsilon = O(1/6)$,
close to the Cabibbo angle that plays a similar role in the construction of quark
mass models. The masses $M_{i}$ remain arbitrary, but leptogenesis is 
possible only if $M_{1} = O(10^{11} \mathrm{GeV})$. Because the $\ell _{i}$
are now given, the present experimental limit on $\mu \rightarrow e\,\gamma$
requires $M_{3} < O(10^{13} \mathrm{GeV}) $, for $\tan \beta = 10$. 
The charges $n _{i}$ are now more or less fixed so that the orders of magnitude of 
all other LFV\&CPV predictions, in particular $d_{e}$, come out much below
the present experiments. With the expected improvement by a factor 30 in 
B.R.$(\mu \rightarrow e\,\gamma)$, one would get $M_{3} < O(10^{11} 
\mathrm{GeV})$  and this very popular  model is close to be excluded. 
What is nice in this exercise is that the constraints from different 
observables apply to 
different parameters and complement each other in an explicit way. 

\section{Conclusions}

The LFV\&CPV experiments, mainly  $\mu \rightarrow e\,\gamma$ and
$d_{e}$ -- but hopefully also $d_{\mu }$  in the near future -- are  a 
selective tool for New Physics model building. They provide indirect tests
which are already sensitive to physics in the TeV region  at the level of 
accuracy of the radiative corrections. They are pure signals of new phenomena
around the TeV scale since the SM contributions are definitely negligible.
It is quite remarkable that they also give  effective tests of the  LFV\&CPV
in seesaw couplings and in grand-unified theories. In particular, the limits on
$d_{e}$ nicely complement the proton decay bounds in selecting O(10)
models! 

The devised improvements in these experiments are important and theoretically 
welcome. The current direct measurement of $d_{\mu }$, improving the present 
indirect bound, motivates a simple question \cite{moha}: to what extent can 
the naive relation $d_{\mu }/d_{e } \simeq  m_{\mu }/m_{e }$ be violated?  
Needless to say, the fact that they are getting into the level of accuracy corresponding 
to where we do expect the new physics to show up means they are on the way
to discover  LFV\&CPV in charged lepton transitions.

\end{document}